\documentclass[pr,twocolumn,showkeys,showpacs,nofootinbib,byrevtex]{revtex4-1}
\usepackage{amsmath,amsfonts,amssymb,amscd,amsxtra,amsthm}
\usepackage{graphicx}
\usepackage{bm}
\usepackage{epstopdf}
\begin{document}

\title{Photoproduction of ${\gamma p\to K^+\Lambda^*(1520)}$ and decay of
$\Lambda^*(1520)\to K^- p$ in the Reggeized framework}
%--------------------------------------------------
\author{Byung-Geel Yu}
\email[E-mail: ]{bgyu@kau.ac.kr} \affiliation{Research Institute
of Basic Sciences, Korea Aerospace University, Goyang, 10540,
Korea}
%--------------------------------------------------
\author{Kook-Jin Kong}
\email[E-mail: ]{kong@kau.ac.kr} \affiliation{Research Institute
of Basic Sciences, Korea Aerospace University, Goyang, 10540,
Korea}
%--------------------------------------------------
%--------------------------------------------------
\date{\today}

\begin{abstract}
Photoproduction of the $\Lambda^*(1520)$ resonance of spin-parity
${3\over2}^-$ off the proton target is investigated within the
Regge framework where the $t$-channel reggeization is applied for
the $K(494)+K^*(892)+K_2^*(1430)$ exchanges in the Born amplitude.
The present model is based on the two basic ingredients; the one
is the minimal gauge prescription for the convergence of the
reaction and the other is the role of the $K_2^*$ crucial to be
consistent with high energy data. The cross sections for the
total, differential and photon polarization asymmetry are
reproduced without fit parameters and compared with existing data.
The LAMP2 and LEPS measurements of the angular distribution of the
$K^-$ in the $\Lambda^*(1520)\to K^-p$ decay are investigated and
found to be dominated by the decay of $\Lambda^*$ with helicity
$\pm3/2$ based on the analysis of the density matrix elements
related. Detailed discussion on the density matrix elements is
given to clarify the analysis of the observable. The reaction
mechanism is featured by the dominance of the contact term with
the $K$ and $K_2^*$ exchanges following in the low energy region.
The $K^*$ exchange appears in minor role. At high energies beyond
$E_\gamma\approx 5$ GeV the role of $K_2^*$ exchange leads over
other exchanges in the reaction process.
\end{abstract}

\pacs{25.20.Lj, % photoproduction reactions
    11.55.Jy, 13.60.Rj, 13.60.Le, 14.40.Df }
\keywords{$\Lambda(1520)$ photoproduction, spin 3/2 resonance,
Regge trajectory, tensor meson exchange, density matrix elements}

\maketitle

\section{introduction}

To study electromagnetic production of strangeness from a nucleon
is important in hadron physics because it provides information
about the interaction between nucleon and hyperon as well as the
static properties of hyperon. There has been a growth of empirical
data dedicated to the study of the hyperons and their resonances
produced by hadronic and electromagnetic probes.

Of these, the reaction $\gamma p\to K^+\Lambda^*(1520)$ is an
interesting process in both sides of  theory and experiment
because the reaction process involves the production of the
negative-parity hyperon-resonance of spin-3/2 via the process
$\gamma p\to K^+K^-p$.

Given the empirical data recently measured with accuracy from the
CLAS \cite{moriya}, LEPS \cite{muramatsu,kohri}, and SAPHIR
\cite{wieland} Collaborations up to photon energy $E_\gamma=3$
GeV, besides the old measurements from the LAMP2 group at higher
energies $E_\gamma= 2.8$ - $4.8$ GeV \cite{barber}, and $11$ GeV
at the SLAC \cite{boyarski}, theoretical efforts on the reaction
process have been made to investigate the production mechanism
based on the effective Lagrangian approach \cite{titov}, or on a
sort of the hybrid model using  the $t$-channel Regge-pole
interpolated with the Feynman propagator \cite{sinam}.

To unravel the role of the baryon resonances in the low energy
region the latter approach was more elaborated to calculate the
contributions of baryon resonances \cite{junhe,xie,wang}.
In these studies the contributions of the $K$ and $K^*$ in the
$t$-channel exchange are analyzed and the role of the $N^*(2120)$
[previously called $D_{13}(2080)$] is identified in the resonance
region.

On the other hand,  it has been an issue to determine the
$K^*N\Lambda^*$ coupling constant from the present process because
it could give a hint on our understanding of  the structure of the
$\Lambda^*$ resonance. In Refs. \cite{roca,hyodo} the $\bar{K}N$
coupled-channel approach to the $\Lambda^*$ leads to the coupling
constant $g_{K^*N\Lambda^*}/m_{K^*}=1.56/m_{K^*}$, whereas the
quark model predicts the value, i.e., $g_{K^*N\Lambda^*}\simeq10$
much larger by an order of magnitude.

In recent experiments it has been another topic to measure the
decay angular distribution of the $K^-$ in the $\Lambda^*(1520)\to
K^-p$ decay process, since the observable is expected to provide
information about spin exchanges of $K$ and $K^*$ in understanding
the production mechanism. Nevertheless, however, the
interpretation of it seems not clear yet, because the recent data
on the photoproduction \cite{wieland} and the result from
electroproduction process \cite{barrow} lead to a conclusion
contradictory to the LAMP2 data.

With these issues in mind, we, here, investigate the $\gamma p\to
K^+\Lambda^*(1520)$ process off the proton target from threshold
to the photon energy 11 GeV, where there is a  data point measured
in the SLAC experiment \cite{boyarski}. Our purpose here is to
provide a theoretical framework which can afford to describe the
reaction without either fit-parameters or any modulation of the
particle propagation for a convergence of the process at high
energies. Indeed, the reaction with the $\Lambda^*$ of spin-3/2 in
the final state would give rise to a divergence as the photon
energy increases.
Nevertheless a special gauge prescription, the so-called the
minimal gauge prescription, adopted in Refs.
\cite{bgyu-pi-delta,bgyu-rho-delta,bgyu-1385} simplifies the
analysis of such  reaction processes to a greater extent. This
interesting result is a consequence of the Ward identity in the
electromagnetic coupling vertex of the spin-3/2 baryon in the
Rarita-Schwinger formalism \cite{stichel}, and supported further
by the agreements with high energy data of $\pi\Delta$ and
$K\Sigma^*$ photoproductions.

We also point out the important role of the spin-2 tensor meson
which is crucial to agree with existing data on $\gamma N\to
K^+\Sigma^*(1385)$ as well as $\gamma p\to \pi^\pm\Delta(1232)$ at
high energies \cite{bgyu-pi-delta,bgyu-1385}. In this work we will
consider the role of the tensor meson $K_2^*$ with an expectation
that the uncertainty in the role of the $K^*$ arisen from the
previous works \cite{titov,sinam,roca,hyodo,toki} but unanswered
yet can be resolved by considering the role of the $K_2^*$ as a
natural parity together with  $K^*$.

This paper is organized as follows: In Sec. II, we provide a
general formalism for the photoproduction amplitude with a brief
introduction of the minimal gauge for the invariant $K$ exchange.
The coupling vertex for the $K_2^*p\Lambda^*$ is newly introduced
with discussion on determining its coupling constants. Numerical
results are given in Section III for the analyses of the old
LAMP2/SLAC data and the recent CLAS/LEPS measurements. This will
includes the cross sections for the total, differential, beam
polarization asymmetry. Analysis of the decay angular distribution
of $K^-$ in the $\Lambda^*\to K^-p$ decay is given in detail based
on the density matrix elements related. We give a summary and
discussion in Section IV. The kinematics related to the
$t$-channel helicity frame of the particle rest is defined in the
Appendix A, and the angular distribution function in terms of the
density matrix elements is given in the Appendix B.

\section{photoproduction amplitude in the minimal gauge}

\begin{figure}[t]{}
\centering
\includegraphics[width=0.65\hsize,angle=0]{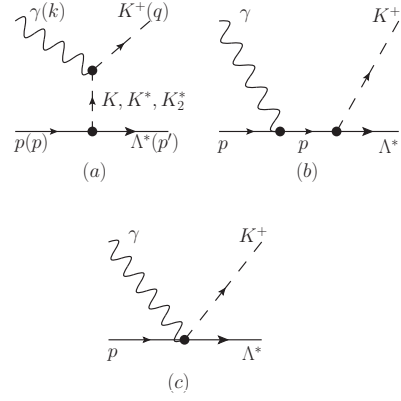}
\caption{Feynman diagrams for $\gamma p \to K^+ \Lambda^{*}$. The
exchange of $K$ in the $t$-channel $(a)$, the proton-pole in the
$s$-channel (b), and the contact term (c) are the basic
ingredients for gauge invariance of the reaction. The $K^*$ and
$K_2^*$ exchanges in the $t$-channel (a) are themselves
gauge-invariant. }\label{fig1}
\end{figure}

We begin with the production amplitude for $\gamma(k)+ p(p)\to
K^+(q)+\Lambda^*(p')$ which is given by
\begin{eqnarray}\label{born}
i{\cal M}=i{\cal M}_{K}+i{\cal M}_{K^*}+i{\cal M}_{K_2^*}\,,
\end{eqnarray}
where each term represents the  $t$-channel $K$, $K^*$, and
$K_2^*$ Regge-pole amplitude in order. The respective 4-momenta of
the photon, proton, kaon, and $\Lambda^*$ are denoted by $k$, $p$,
$q$, and $p'$ in Fig. \ref{fig1} (a).

In these terms the latter two exchanges are themselves gauge
invariant. However, the $K$ exchange in the $t$-channel is not
gauge invariant as expressed in the Born amplitude, i.e.,
\begin{eqnarray}\label{kaon}
i{M}_{t(K)}=e_K\frac{f_{KN\Lambda^*}}{m_K}\bar{u}_{\Lambda^*}^\nu(p')\gamma_5Q_\nu
\frac{2q\cdot\epsilon}{t-m_K^2}u_N(p),
\end{eqnarray}
where $e_K$ is the kaon charge, $\epsilon$ is photon polarization,
and $Q^\nu=(q-k)^\nu$ is the $t$-channel momentum transfer.

%%%%%%%%%%%%%%%%%%%%%  FIG 1 %%%%%%%%%%%%%%%%%%%%%%%%%%%
%%%%%%%%%%%%%%%%%%%%%%%%%%%%%%%%%%%%%%%%%%%%%%%%%%%%%%%%

For the gauge invariant $K$ exchange in Eq. (\ref{kaon}) the
current conservation following the charge conservation,
$e_p-e_{K}-e_{\Lambda^*}=0$, requires the inclusion of the charge
term of the proton-pole in the $s$-channel and the contact term.
Here, $e_p$ and $e_{\Lambda^*}$ are the respective charges of
proton and $\Lambda^*$.
Since  the usual Dirac charge-coupling term in the proton-pole is,
$e_p(\rlap{/}p+\rlap{/}k+M)\rlap{/}\epsilon=e_p(2p\cdot\epsilon+\rlap{/}k\rlap{/}\epsilon)$,
it can be reduced to $e_p\,2p\cdot \epsilon$ by removing the
transverse term by redundancy with respect to gauge invariance in
the sense of the minimal \footnote{ The minimal gauge requires the
pure charge term of the proton-pole to be included. If one uses
$e_p(\rlap{/}p+\rlap{/}k+M_p) \rlap{/}\epsilon$ in Eq.
(\ref{proton}), but still containing the transverse component,
he(she) cannot obtain an agreement with the cross section data
without a cutoff function as shown in Fig. \ref{fig2} (a) below. }
as discussed in Refs. \cite{bgyu-pi-delta,bgyu-1385}.

Thus, the gauge-invariant extension of the $K$ exchange which is
now Reggeized is given by
\begin{eqnarray}\label{eq3}
&&i{\cal M}_{K}=i\left[{ M}_{t(K)}+{M}_{s(p)}+{
M}_{c}\right]\times(t-m_K^2)\nonumber\\&&\hspace{1cm}\times{\cal
R}^{K}(s,t)e^{-i\pi\alpha_K(t)},
\end{eqnarray}
where
\begin{eqnarray}\label{proton}
&&i{M}_{s(p)}=e_p\frac{f_{KN\Lambda^*}}{m_K}
\bar{u}_{\Lambda^*}^{\nu}(p')\gamma_5 q_{\nu}\frac{2p\cdot
\epsilon}{s-M^{2}_{p}}u_N(p)\,,\\
&&i{M}_c=-e_K\frac{f_{KN\Lambda^*}}{m_K}\,
\bar{u}_{\Lambda^*}^\nu(p')\gamma_5\epsilon_\nu
u_N(p)\,,\label{contact}
\end{eqnarray}
are the  proton-pole and the contact term, and
\begin{eqnarray}
{\cal
R}^\varphi(s,t)={\pi\alpha_\varphi'\over\Gamma[\alpha_\varphi(t)+1-J]
\sin\pi\alpha_\varphi(t)}\left({s\over
s_0}\right)^{\alpha_\varphi(t)-J}\,,
\end{eqnarray}
is the Regge propagator written collectively for the
$\varphi(=K,\,K^*,\,K_2^*)$ of spin-$J$. The $\alpha_\varphi(t)$
is the trajectory and $s_0=1$ GeV$^2$ is taken as usual.

For the higher-spin meson exchanges we consider the vector-meson
$K^*$ and the Reggeized amplitude relevant to the leading
contribution of the $K^*N\Lambda^*$ interaction is given by
\begin{eqnarray}\label{kstar}
&&i\mathcal{M}_{K^*}=-i{g_{\gamma KK^*}\over
m_0}\frac{f_{K^*N\Lambda^*}}{m_{K^*}}
\,\epsilon^{\alpha\beta\lambda\sigma}\epsilon_{\alpha}
k_{\beta}q_{\lambda}\nonumber\\
&&\hspace{0.5cm}\times\bar{u}^{\nu}(p')\left(\rlap{/}{Q}
g_{\nu\sigma}-Q_{\nu}\gamma_\sigma \right)u(p){\cal
R}^{K^*}(s,t)e^{-i\pi\alpha_{K^*}(t)},\ \ \
\end{eqnarray}
with the mass parameter $m_0=1$ GeV.

In accordance with our previous work on $\gamma N\to
K^+\Sigma^*(1385)$, we introduce the tensor meson $K_2^*$ exchange
in the $t$-channel to the present reaction process. By considering
the parity of the $\Lambda^*$ opposite to the $\Sigma^*$ the new
Lagrangian for the $K_2^*N\Lambda^*$ coupling firstly considered
in this work can be written as
\begin{eqnarray}
{\cal L}_{K^*_2N\Lambda^*}=i{f_{K_2^*N\Lambda^*}\over
m_{K_2^*}}\bar{~\Lambda^*}_\alpha\left(g^{\alpha\mu}\partial^\nu
+g^{\alpha\nu}\partial^\mu\right) N {K_2^*}_{\mu\nu}\,,
\end{eqnarray}
and the Reggeized amplitude for the $K_2^*$ exchange is given by
\begin{eqnarray}\label{k2star}
&&i\mathcal{M}_{K_2^*}=-i\frac{2g_{\gamma
KK_2^*}}{m_0^2}{f_{K_2^*N\Lambda^*}\over m_{K_2^*}}
\,\epsilon^{\alpha\beta\mu\lambda}\epsilon_\mu k_\lambda Q_\alpha
q_\rho
\Pi^{\beta\rho;\sigma\xi}(Q)\nonumber\\&&\hspace{0.5cm}\times
\bar{u}^{\nu}(p')(g_{\nu\sigma}P_\xi+g_{\nu\xi}P_\sigma)u(p){\cal
R}^{K^*_2}(s,t)e^{-i\pi\alpha_{K_2^*}(t)},
\end{eqnarray}
with the momentum  $P=(p+p')/2$, and
$\Pi^{\beta\rho;\sigma\xi}(q)$ the spin-2 projection given in Ref.
\cite{bgyu-1385}.
\\

The coupling constant $f_{KN\Lambda^*}$ is estimated to be
$f_{KN\Lambda^*}=10.59$ from the measured decay width
$\Gamma_{\Lambda^*\to \bar{N}K}=7.02$ MeV in the Particle Data
Group.

The radiative decay constant in the  $K^*$ exchange is estimated
to be  $g_{\gamma K^+K^{*+}}=\pm0.254$ from the measured widths
$\Gamma_{K^*\to\gamma K^\pm }=(0.05\pm0.005)$ MeV, and we take the
negative sign to agree with empirical data. As mentioned before,
however the determination of the coupling constant
$f_{K^*N\Lambda^*}$ is still at large.
We found  in Ref. \cite{titov} that  the phenomenological
determination of the $K^*N\Lambda^*$ coupling constant was
discussed rather in detail, in which case
$g_{K^*N\Lambda^*}=\alpha_{\Lambda^*}g_{KN\Lambda^*}$ was assumed
with the parameter $\alpha_{\Lambda^*}=+0.372$, or $-0.657$
obtained from the  fit of the LAMP2 data \cite{titov}. By the
different  definition of the coupling vertex in the present work
these correspond to $f_{K^*N\Lambda^*}=+7.12$ and $-12.57$,
respectively. But we note that the values of the
$\alpha_{\Lambda^*}$ were determined by using the $K^*$-trajectory
$\alpha_{K^*}(t)=0.9\,t-0.1$ and the energy-dependence of the
Regge-pole $\sim s^{\alpha_{K^*}(t)}$, %apart from the constant $C=29.6$,
which are quite different from those we have employed in the
present calculation. Moreover, such a lower intercept of the
trajectory significantly reduces the contribution of the
Regge-pole exchange so that a large coupling constant might be
needed in their analysis in order for an agreement with the LAMP2
data.

%%%%%%%%%%%%%%%%%%%  tb1  %%%%%%%%%%%%%%%%%%%%%%%%%%%%%%%%
%\begin{table*}
\begin{table}
\caption{Coupling constants of exchanged mesons in
$\Lambda^*(1520)$ process. The meson-baryon coupling constants in
the set I are for the CLAS and LEPS data and the set II for the
LAPM2 and SLAC data. The signs of radiate decay constants
$g_{\gamma KK^*}=-0.254$ and $g_{\gamma KK^*_2}=-0.276$ are taken
in this work. The quark model prediction for
$f_{KN\Lambda^*}\simeq 10$.}
\begin{ruledtabular}\label{tb1}
\begin{tabular}{ccc|c|c}
&   Refs. \cite{roca,hyodo} &  Ref. \cite{titov} &Set I & Set II\\
\hline
$f_{KN\Lambda^*}/\sqrt{4\pi}$&$-0.15$ & 2.99 &1.8      &  2.5       \\%
\hline
$f_{K^*N\Lambda^*}$    &$\pm1.58$&$-12.57$, $7.12$&$3.5$ & 3.5\\%
\hline
${f_{K^*_2N\Lambda^*}\over m_{K_2^*}}$  &  -  &  - & $-{10.5\over m_{K^*}}$&$-{10.5\over m_{K^*}}$\\%
\end{tabular}
\end{ruledtabular}
\end{table}
%%%%%%%%%%%%%%%%%%%%%%%%%%%%%%%%%%%%%%%%%%%%%%%%%%%%%%%%%%%%

In this work we consider to determine the $f_{K^*N\Lambda^*}$ from
the numerical analysis of the reaction.  Within the present
framework, however, it is not a free parameter because it is
related with coupling constant of the tensor meson $K_2^*$ as
\begin{eqnarray}\label{symm}
{f_{K_2^*N\Lambda^*}\over m_{K_2^*}}=-3{f_{K^*N\Lambda^*}\over
m_{K^*}}
\end{eqnarray}
by the duality and vector dominance  \cite{goldstein}. Since the
latter exchange is expected to play the role significant in the
high energy region, we first obtain the $K_2^*$ coupling constant
as $-3f_{K^*N\Lambda^*}/m_{K^*}=-10.5/m_{K^*}$ by fitting to the
SLAC data point at $E_\gamma=11$ GeV, and then, the
$f_{K^*N\Lambda^*}$ is determined according to the relation in Eq.
(\ref{symm}) above. We adopt these coupling constants in the case
of spin-3/2 baryon of negative parity, because the duality and the
vector dominance are believed to be universal.
The radiative coupling constant for the $K_2^*$ is estimated to be
$g_{\gamma KK_2^*}=\pm0.276$ from the decay width
$\Gamma_{K_2^*\to\gamma K^\pm }=(0.24\pm0.05)$ MeV
\cite{bgyu-kaon}. In the calculation we choose the negative sign
for a better agreement with existing data.

The coupling constants for the present calculation are summarized
in Table \ref{tb1} and compared to those used for other model
calculations.

As to the trajectory and phase of the $t$-channel Regge pole, we
use
\begin{eqnarray}\label{regge2}
&&\alpha_K(t)=0.7\,(t-m_K^2)\,,\nonumber\\
&&\alpha_{K^*}(t)=0.83\,t+0.25\,,\nonumber\\
&&\alpha_{K^*_2}(t)=0.83\,(t-m_{K^*_2}^2)+2\,,%\nonumber\\
\end{eqnarray}
for a consistency with the previous works
\cite{bgyu-1385,bgyu-kaon,bgyu-kstar}, and assign the complex
phase to  all the mesons $K$, $K^*$, and $K^*_2$ as shown above.

\section{numerical results}

In this section  we first discuss our result in the differential
cross sections from the LAMP2 and SLAC experiment to examine the
convergence of the reaction at high energies. Then, we present the
cross sections from the CLAS and LEPS data in the lower energy
region with the relevance of the coupling constants to the old and
recent measurements indicated in Table \ref{tb1}.

In order to avoid confusion in expression in what follows, we
denote the symbol $K$ by the single $K$ exchange in Eq.
(\ref{kaon}) which is Reggeized by multiplying $(t-m^2_K){\cal
R}^Ke^{-i\pi\alpha_K}$. Similarly, the symbols $K^*$ and $K_2^*$
denote the amplitudes in Eqs. (\ref{kstar}) and (\ref{k2star}) for
brevity. Also we call the contact term to imply the amplitude in
Eq. (\ref{eq3}) without $M_{t(K)}$ and $M_{s(p)}$.

\subsection{Reaction at high energies}

\begin{figure}[]
%\vspace{.5cm}
\includegraphics[width=8.1cm]{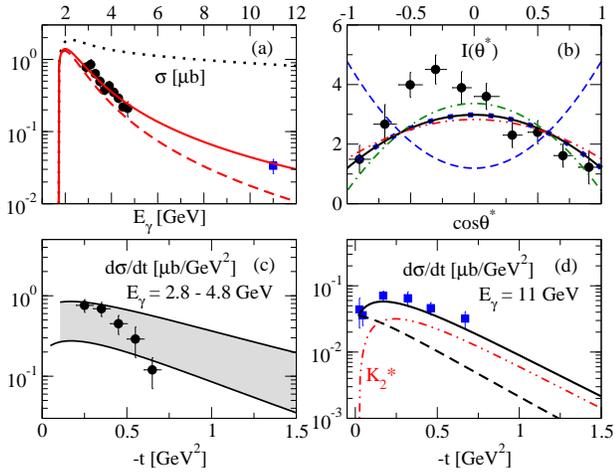}
\caption{Total and differential cross sections in (a) and (c) at
$E_\gamma$=2.8 $\sim$ 4.8 GeV  \cite{barber}, and the angular
distribution of $K^-$ in the decay $\Lambda^*\to K^-p$ in (b) from
the LAMP2 measurement.  The differential cross section from the
SLAC at $E_\gamma=11$ GeV is shown in (d) \cite{boyarski}.
The dotted line in (a) results from the proton-pole with the
charge term, $e_p(\rlap{/}p+\rlap{/}k+M_p)\rlap{/}\epsilon$ in Eq.
(\ref{proton}) for the full amplitude in Eq. (\ref{born}). The
cross sections without $K_2^*$ are shown by the dashed lines in
the total (a) and differential cross sections (d). In (b) the
decay angle of $K^-$ is denoted by $\theta^*$ for a distinction
from the reaction angle $\theta$ of the $K^+$ produced. The
$I(\theta^*)$ is reproduced at $E_\gamma=3.8$ GeV and
$\theta=1^\circ$ by the density matrix elements $\rho_{33}$ and
$\rho_{11}$ in Fig. \ref{fig3} with the factor of 30 multiplied to
Eq. (\ref{itheta}). The contributions of $K$, $K^*$, and $K_2^*$
are depicted by the blue dashed, green dash-dotted, red
dash-dot-dotted line, in order. The blue dotted line is from the
contact term and exactly coincides with the solid line from the
full calculation. Note that the contribution of the $K$ exchange
from Eq. (\ref{kaon}) is zero at $\theta=0^\circ$ in (b).}
\label{fig2}
\end{figure}

Before the LEPS and CLAS experiments \cite{muramatsu,kohri,moriya}
theoretical studies \cite{titov,toki} on this reaction had been
based on the LAMP2 and SLAC data which were shown in Fig.
\ref{fig2} for the total, differential and the angular
distribution of $K^-$ in the decay $\Lambda^*\to K^-p$ in the
final state.

%%%%%%%%%%%%%%%%% FIG %%%%%%%%%%%%%%%%%%%%%%%%%%%%%
%%%%%%%%%%%%%%%%%%%%%%%%%%%%%%%%%%%%%%%%%%%%%%%%%%%%%

In Fig. \ref{fig2} the LAMP2 data for the total and differential
cross sections at $E_\gamma=2.8\sim4.8$ GeV and the SLAC
differential cross section at $E_\gamma=11$ GeV are reproduced by
the full amplitude in Eq. (\ref{born}) with the coupling constants
from the set II in Table \ref{tb1}.
The leading coupling constant $f_{KN\Lambda^*}$ taken here is
smaller than the value estimated from the measured decay width  by
a factor of $0.84$. In order to test the validity of the minimal
gauge we reproduce the total cross section with the charge term
taken as $e_p(\rlap{/}p+\rlap{/}k+M_p)\rlap{/}\epsilon$ in the
proton-pole  instead of $e_p\,2p\cdot\epsilon$ in Eq.
(\ref{proton}) and show the result by the dotted line in (a). This
demonstrates that such a good convergence of the cross section to
the experimental data cannot be obtained without the cutoff
function, otherwise. Also it should be noted that the role of the
$K_2^*$ is crucial to reproduce the total and differential cross
sections at high energy, as can be seen by comparing the solid
lines with the dashed ones in (a) and (d).

The final state $\Lambda^*(1520)$ of spin-parity $3/2^-$ decaying
to $K^- p$ system was measured in the LAMP2 group where the
angular distribution of the $K^-$ is given by
\begin{eqnarray}\label{itheta}
I(\theta^*)={3\over4\pi}\left[\rho_{33}\sin^2\theta^*+\rho_{11}\left({1\over3}
+\cos^2\theta^* \right) \right],
\end{eqnarray}
with the angle $\theta^*$ of the decaying $K^-$  in the
$\Lambda^*$ rest frame. Here, we neglect the remnants in the
original equation in Ref. \cite{barber}  by fixing the angle
$\phi^*=90^\circ$ as well as the smallness of the density matrix
${\bf\rm Re} [\rho_{3-1}]$ for simplicity.
Since the dynamics of the photoproduction process is contained in
the density matrix elements $\rho_{33}$ and $\rho_{11}$, their
dependences on the energy and angle are important to  analyze the
$I(\theta^*)$ in conjunction with the reaction mechanism of
photoproduction.

%%%%%%%%%%%%%%%%% FIG %%%%%%%%%%%%%%%%%%%%%%%%%%%%%
\begin{figure}[]
%\bigskip
%\bigskip
\includegraphics[width=8cm]{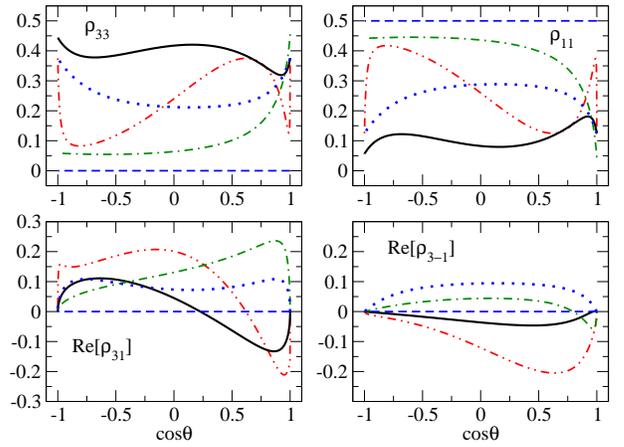}
\caption{Density matrix elements for the $\gamma p\to
K^+\Lambda^*$ at $E_\gamma=3.8$ GeV. The angle $\theta$ dependence
of the $\rho_{\lambda\lambda'}$ is calculated in the $t$-channel
helicity frame of $\Lambda^*$ \cite{muramatsu,barrow,schilling}.
Solid lines are the full calculation of the
$\rho_{\lambda\lambda'}$ from the set II with the conventions and
definitions from Refs. \cite{donnachie}. The respective
contributions of $K$, $K^*$, and $K_2^*$ are shown by the blue
dashed, green dash-dotted, red dash-dot-dotted curves in order.
The blue dotted line is from the contact term. The density matrix
elements unravel the overall dependence of the $I(\theta^*)$ on
the angle. At the $\theta=0^\circ$, $\rho_{33}=0.375$ and
$\rho_{11}=0.125$, respectively, to satisfy the trace condition
$\rho_{33}+\rho_{11}=1/2$.  } \label{fig3}
\end{figure}
%%%%%%%%%%%%%%%%%%%%%%%%%%%%%%%%%%%%%%%%%%%%%%%%%%%%%

In the photoproduction process associated with the decay channel,
$\gamma p\to K^+\Lambda^*\to K^+K^-p$, the LAMP2 data was
understood as the dominance of the $\Lambda^*$ decay with the
helicity $S_z=\pm3/2$,
%
%the natural parity exchange \cite{barber},
as shown in Fig. \ref{fig2} (b) where  the $K^*$ exchange of the
natural parity in the photoproduction leads to the decay of the
$\Lambda^*$ with helicity $S_z=\pm{3\over2}$, which corresponds to
the term of $\sin^2\theta^*$, whereas the $K$ exchange of
unnatural parity to the $S_z=\pm{1\over2}$ corresponding to the
term of $\left({1\over3}+\cos^2\theta^*\right)$.

Figure \ref{fig3} shows the density matrix elements $\rho_{33}$
and $\rho_{11}$ calculated in the $t$-channel helicity frame of
$\Lambda^*$ where the $z$-axis is taken to be the direction
opposite to the target proton momentum
\cite{muramatsu,barrow,schilling}, as presented in the Appendix.
At the $\theta\approx0^\circ$ and at $E_\gamma=3.8$ GeV, together
with the small negative values for the Re$[\rho_{32}]$ and
Re$[\rho_{3-1}]$, we obtain $\rho_{33}=0.375$ and
$\rho_{11}=0.125$, which are consistent with those extracted from
the LAMP2 data given in Table 1 of Ref. \cite{barber}.
As the representative of each term in Eq. (\ref{itheta}),
therefore, the angular dependence of the $\rho_{\lambda\lambda'}$
in Fig. \ref{fig3} can tell us that the $\rho_{33}$ is always
dominant as compared to the $\rho_{11}$ in the overall range of
the angle at the given energy. Thus, we expect the dominance of
the natural parity exchange in the photoproduction from the shape
of the $I(\theta^*)$ convex up and this is consistent with the
LAMP2 data as we reproduced in Fig. \ref{fig2} (b).

Within the present framework which is valid for the forward angles
the dominance of the $\sin^2\theta$ term is due to the
contribution of the contact term which dominates the
photoproduction below $E_{\gamma}\approx 4$ GeV as shown in Fig.
\ref{fig4} in addition to the contributions of the $K^*+K_2^*$ at
the very forward angle. As can be seen in Fig. \ref{fig3},
however, the contribution of the $K^*_2$ is varying very rapidly
near $\theta=0^\circ$ in shaping the $I(\theta^*)$. Nevertheless,
we find that the $K$ exchange always yields the shape concave down
%by the continuing contribution to the $\left(1/3+\cos^2\theta\right)$ term
by the role solely in the $\rho_{11}$.

Differential cross sections at $E_\gamma=3.8\pm1$ GeV from the
LAPM2 and 11 GeV from the SLAC experiments are presented in Fig.
\ref{fig3} (c) and (d). As shown by the dash-dot-dotted line in
the latter case,  the role of the $K^*_2$ is apparent, and the
dashed line in (d) as well as in (a)  without it cannot agree with
the experimental data for the differential and the total cross
sections.

\subsection{Reaction at low and in the intermediate energy}

%%%%%%%%%%%%%%%%% FIG %%%%%%%%%%%%%%%%%%%%%%%%%%%%%
\begin{figure}[]
\includegraphics[width=8cm]{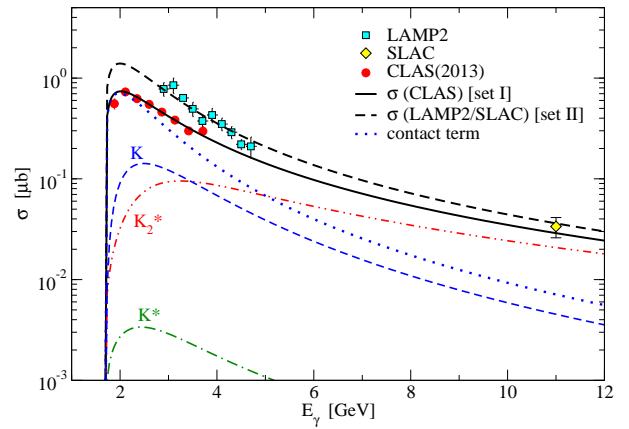}
\caption{Total cross sections for $\gamma p\to
K^+\Lambda^*(1520)$. Solid line is the full calculation of the
cross section from the CLAS data with coupling constants set I in
Table \ref{tb1}. The dashed line is the total cross section of the
LAMP2 and SLAC data with coupling constants of set II. The
contributions of the contact term and the meson exchanges are
shown with the same notations as in Fig. \ref{fig3}. The CLAS Data
are taken from Ref. \cite{moriya}. } \label{fig4}
\end{figure}
%%%%%%%%%%%%%%%%%%%%%%%%%%%%%%%%%%%%%%%%%%%%%%%%%%%%%

%%%%%%%%%%%%%%%%% FIG %%%%%%%%%%%%%%%%%%%%%%%%%%%%%
\begin{figure}[]
\bigskip
\bigskip
\includegraphics[width=8cm]{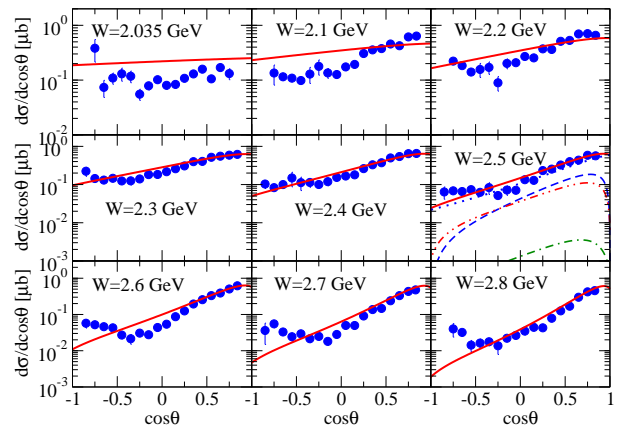}
\caption{Differential cross sections of the CLAS Collaboration in
the nine energy bins.  The solid line results from physical
constants of the set I. The leading role of the contact term with
the $K$ and $K_2^*$ following with equal contributions are well
reproduced in the differential cross sections as well as the total
cross section above. The contribution of meson exchange and the
contact term are displayed with the same notations as in Fig.
\ref{fig4}. Data are taken from Ref. \cite{moriya}.} \label{fig5}
\end{figure}
%%%%%%%%%%%%%%%%%%%%%%%%%%%%%%%%%%%%%%%%%%%%%%%%%%%%%

%%%%%%%%%%%%%%%%% FIG %%%%%%%%%%%%%%%%%%%%%%%%%%%%%
\begin{figure}[]
%\bigskip
%\bigskip
\includegraphics[width=8cm]{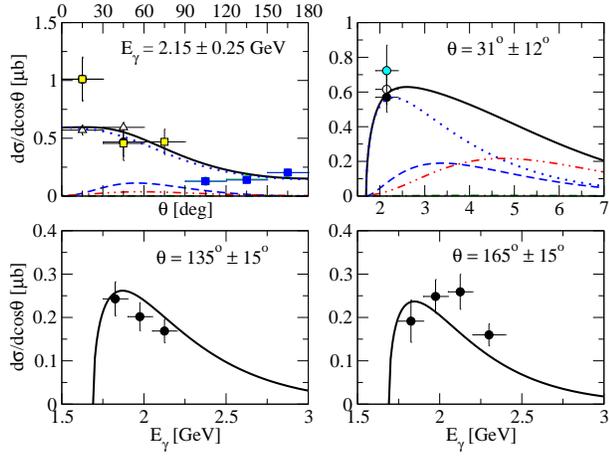}
\caption{Angle and energy dependences of differential cross
sections measured in the range of the energy $1.9 < E_\gamma <
2.4$ GeV and the reaction angle $30<\theta<180$. The solid lines
from the set I show the results at the central values in the
experimental ranges. The contribution of meson exchange and the
contact term are displayed with the same notations as in Fig.
\ref{fig4}. Data of the LEPS in 2009 (black filled circle) are
taken from Ref. \cite{muramatsu}.} \label{fig6}
\end{figure}
%%%%%%%%%%%%%%%%%%%%%%%%%%%%%%%%%%%%%%%%%%%%%%%%%%%%%

We now turn to the analysis of the experimental data recently
measured by the CLAS and LEPS Collaborations.

Figure \ref{fig4} shows the total cross section from the set I in
Table \ref{tb1} and the result is compared with the CLAS data. The
dashed line of the LAMP2 cross section in Fig. \ref{fig2} is also
reproduced  for comparison. Note that the discrepancy between the
two, and we use the $f_{KN\Lambda^*}/\sqrt{4\pi}=1.8$ in order to
agree with new data. The dominance of  the contact term over the
meson exchanges in the CLAS data is illustrated. It should be
pointed out that the exchange of the $K_2^*$ as the natural parity
becomes the leading contribution at high energy $E_\gamma=11$ GeV
to meet with the SLAC data.
In contrast, the role of the $K^*$ exchange is very small in
comparison to those of $K$ and $K^*_2$ as shown in Fig.
\ref{fig4}. This may explain why the cross sections is not
sensitive to such a wide change of $K^*N\Lambda^*$ from $0$ to
$\pm11$ in the model without the tensor meson $K_2^*$
\cite{sinam}.

Together with the total cross section in Fig. \ref{fig4} the
differential cross sections displayed in Fig. \ref{fig5} shows
that the reaction mechanism is feature by the leading role of the
contact term followed by the $K$ and $K_2^*$ with equal
contributions, and this feature from the CLAS measurement is also
valid for the differential cross sections measured in the LEPS
experiment as will be shown in Fig. \ref{fig6}  next.

%%%%%%%%%%%%%%%%% FIG %%%%%%%%%%%%%%%%%%%%%%%%%%%%%
\begin{figure}[]
\bigskip
\includegraphics[width=8.3cm]{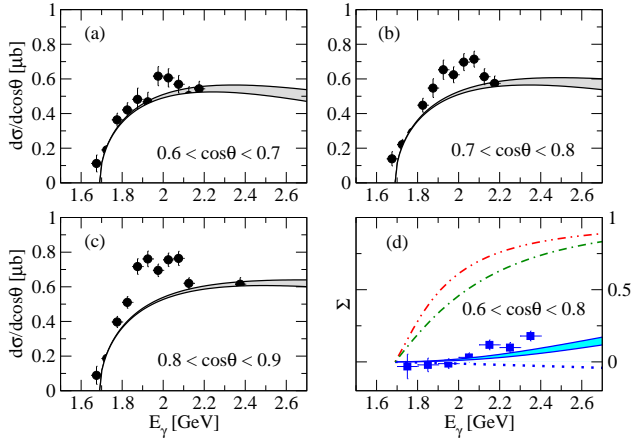}
\caption{Energy dependence of differential cross sections and beam
polarization asymmetry measured in angle bins. The bands
correspond to the cross sections in the range of angles denoted in
each figure with coupling constants from the set I. The
contribution of meson exchange and the contact term are displayed
in the $\Sigma$ with the same notations as in Fig. \ref{fig4}.
Data of the LEPS in 2010 are taken from Ref. \cite{kohri}.}
\label{fig7}
\end{figure}
%%%%%%%%%%%%%%%%%%%%%%%%%%%%%%%%%%%%%%%%%%%%%%%%%%%%%

Shown in Figs. \ref{fig6} and \ref{fig7} are the energy and angle
dependences of the differential cross sections and the energy
dependence of the beam polarization asymmetry measured in a
sequential experiment of the LEPS Collaboration.
The roles of the contact term and each meson exchange discussed in
the CLAS data are  realized in these differential cross sections
as well with respect to energy and angle, as shown in Fig.
\ref{fig6}. Therefore, such a  consistency of the LEPS data with
the CLAS confirms the validity of the present analysis based on
the Reggeized framework without fit parameters.
However, the lack of the present model predictions in the backward
region of the CLAS data in Fig. \ref{fig5} as well as in the
threshold peaks in the LEPS as shown in Fig. \ref{fig7} (a), (b),
(c) is suggestive of the contributions from the baryon resonances
to the reaction process. The LEPS data in 2009 \cite{muramatsu}
reported that $\Sigma\simeq -0.01\pm0.07$ at the kaon angle
$\theta$ less than $60^\circ$ in the energy interval
$E_\gamma=1.75\sim 2.4$ GeV, and suggested that it is almost zero
within the experimental uncertainties. The data from the LEPS in
2010 \cite{kohri} showed a slightly positive value of $\Sigma$ in
the same energy range but for the different angles $0.6<
\cos\theta < 1$, and we reproduce it in the range of the angle
$0.6\leq\cos\theta\leq0.8$ by using
$\Sigma={d\sigma_y-d\sigma_x\over d\sigma_y+d\sigma_x}$, where $x$
and $y$ are the axes of the reaction plane perpendicular to the
$z$-axis taken to be the incident photon momentum, as usual. The
result presented in Fig. \ref{fig7} (d) shows the large positive
contributions from the $K^*+K_2^*$ of the natural parity and the
large negative contribution of the $K$ exchange, i.e., $-1$, to
the $\Sigma$, though not shown here. Hence the difference between
the natural and unnatural parity leads to the positiveness of the
$\Sigma$.

%%%%%%%%%%%%%%%%% FIG %%%%%%%%%%%%%%%%%%%%%%%%%%%%%
\begin{figure}[]
%\bigskip
\includegraphics[width=8cm]{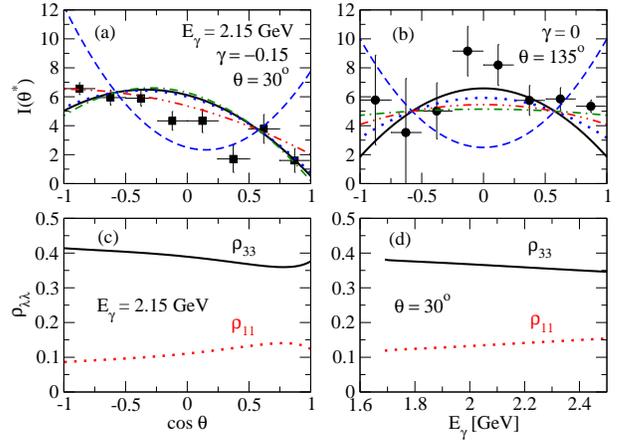}
\caption{Angular distributions of the polar $K^-$ in the
$t$-channel helicity frame of $\Lambda^*$. Solid curves are our
results in the $I(\theta^*)$ from the set I with the parameters
$\alpha$ and $\beta$ replaced by $\rho_{33}$ and $\rho_{11}$ in
Eq. (\ref{itheta2}). The solid curves in the forward (a) and
backward (b) directions are estimated at $\theta=30^\circ$ with
$\gamma=-0.15$, and at $\theta=135^\circ$ with $\gamma=0$ for the
fixed $N=15$ at $E_{\gamma}=2.15$ GeV. The contributions of the
$K$, $K^*$, $K_2^*$, and the contact term are shown with the same
notations as in Fig. \ref{fig3}.  Data are taken from Ref.
\cite{muramatsu}.
The angle and energy dependences of the $\rho_{33}$ and
$\rho_{11}$ at forward angle (a) are presented by solid and dotted
lines in (c) and (d). } \label{fig8}
\end{figure}
%%%%%%%%%%%%%%%%%%%%%%%%%%%%%%%%%%%%%%%%%%%%%%%%%%%%%

Finally, we reexamine the $K^-$ angular distribution $I(\theta^*)$
in the  $\Lambda^*$ rest frame by using the LEPS measurement. The
data were obtained by using a fit of $K^+p$ mode based on the
function,
\begin{eqnarray}\label{itheta2}
I(\theta^*)=N\left[\alpha\sin^2\theta^*
+\beta\left({1\over3}+\cos^2\theta^*\right)
+\gamma\cos\theta^*\right],\nonumber\\
\end{eqnarray}
where the effect of the background is further assumed in the last
term with the parameter $\gamma$ and a scale constant $N$ to the
arbitrary dimension of $I(\theta^*)$. The fractions of the
$\Lambda^*$ helicities $\pm3/2$  and $\pm1/2$ are parameterized as
$\alpha$ and $\beta$, which were extracted to be about 0.52 and
0.48 from the LEPS fitting procedure in the kinematical ranges
$1.9 <E_\gamma < 2.4$ GeV and $0^\circ<\theta<60^\circ$. In the
backward region the fractions  $\alpha\approx0.63$ and
$\beta\approx 0.37$ were extracted  in the angle $90^\circ<\theta<
180^\circ$ and energy range $1.7<E_\gamma<2.4$ GeV.
%
%Thus, the $I(\theta^*)$ is completely controlled by the parameters
%$N$ and $\beta$ for its size and tilt in the fit of data. But,
%
Here, we consider to recover the $\alpha$ and $\beta$ as the
$\rho_{33}$ and $\rho_{11}$, as before, in order to give them the
energy and angle dependences in Eq. (\ref{itheta2}). Given the
constant $N=15$, we obtain those solid curves in Fig. \ref{fig8}
with $\gamma=-0.15$ at $\theta=30^\circ$ for the forward (a) and
$\gamma=0$ at $\theta=135^\circ$  for the backward (b) directions
at $E_\gamma=2.15$ GeV by using the $\rho_{33}$ and $\rho_{11}$
calculated in the $t$-channel helicity frame at the given energy
and angles.

In order to examine the variation of the  $I(\theta^*)$ at the
central values of angle and energy as in (a) and (b) within the
experimental ranges of angle and energy, we calculate the angle
and energy  dependences of the $\rho_{33}$ and $\rho_{11}$  to
confirm that the variations of these variables are not significant
as shown in (c) and (d). Furthermore,  the overall dominance of
the $\rho_{33}$ over the $\rho_{11}$ is apparent in the given
ranges of angle and energy. Thus, we compare
$\rho_{33}\approx0.38$ and $\rho_{11}\approx0.12$ to those
$\alpha/2\approx 0.26$ and $\beta/2\approx0.24$ from the fit of
the LEPS above at forward angle, by considering that the trace
condition of the latter is twice that of the former.  This
supports the result in the analysis of the LAMP2 data.

\section{summary and discussion}

In this work we have investigated the reaction $\gamma p\to
K^+\Lambda^*(1520)$ from threshold to photon energy $E_\gamma=11$
GeV based on the production amplitude in Eq. (\ref{born}) for the
meson exchanges in the $t$-channel. Following the convention and
definitions of the previous works
\cite{bgyu-kaon,bgyu-kstar,bgyu-1385} the $t$-channel exchanges
are Reggeized with the trajectories taken the same as those
$\gamma p\to K^+\Lambda$, $\gamma p\to K^{*+}\Lambda$, and $\gamma
p\to K^+\Sigma^*(1385)$ for consistency's sake.

Covering whole range of the reaction energy for the  CLAS, LEPS ,
LAMP2 and SLAC experiments the cross sections for  total,
differential, and beam polarization asymmetry are analyzed without
either cutoff functions or fit-parameters. The angular
distribution of the $K^-$ from the $\Lambda^*\to K^-p$ decay in
the final state is discussed based on the role of the density
matrix elements played in the LAMP2 and LEPS data.
The basic ingredients of this simple model are the minimal gauge
prescription for the convergence of the reaction to the
experimental data and the role of the $K_2^*$ substantial to agree
with data at high energies.
Within the present framework, therefore, the electromagnetic
production of spin-3/2 baryon resonance of negative parity could
simply  be understood as the production mechanism similar to the
well-known cases of $\pi\Delta$ \cite{bgyu-pi-delta} and
$K\Sigma^*$ \cite{bgyu-1385}, just as we have illustrated here.

Through out the analysis of the total and differential cross
sections we find that the production mechanism is featured by the
dominance of the contact term and the $K$ and $K_2^*$ exchanges
follow with almost equal contributions in the low energy region.
While the role of the $K^*$ exchange is minor, the $K_2^*$
exchange plays the role leading over other exchanges in the
reaction process at high energies.
Though determined from the phenomenological analysis, here, the
coupling constant of $f_{K^*N\Lambda^*}=3.5$ is interesting
because it is closer to the result of the $\bar{K}N$ coupled
channel approach \cite{hyodo} rather than the quark model
prediction. It should be recalled that the determination of the
$K^*$ in this work is not a free parameter but bound to a
determination of the $K_2^*$ coupling constant by Eq.
(\ref{symm}), which is set to agree with the SLAC data at
$E_\gamma=11$ GeV.  Therefore,   the difficulty in identifying the
coupling strength of the $K^*$ in previous studies is resolved by
including the $K_2^*$ with both coupling constants set to be
determined simultaneously in one relation.

Based on the density matrix elements presented here,  we analyze
the angular distribution function $I(\theta^*)$ for the
$\Lambda^*\to K^-p$ decay to investigate the role of the natural
and unnatural parity exchanges in the photoproduction process.  By
using the $t$-channel helicity frame of $\Lambda^*$ for the
calculation of the density matrix elements
$\rho_{\lambda\lambda'}$, we show that the numerical analysis of
the LAMP2 and LEPS data on  the $I(\theta^*)$ is consistent with
the dominance of the helicity $\pm3/2$ of $\Lambda^*$ decay due to
the strong contribution of the $\rho_{33}$ by the contact term in
addition to the $K^*+K_2^*$ contributions. This finding is,
however, contradictory to the experiments in the SAPHIR
\cite{wieland} and the CLAS albeit the latter case of
electroproduction has more reasons to be \cite{barrow}. Therefore,
such a contradiction in explaining the $I(\theta^*)$ should be
clarified in experiments by the precise measurements of the
$\rho_{33}$ and $\rho_{11}$ in a specified G.-J. frame.

As an application of the present work it is desirable to reexamine
the $N^*$ resonances here because the role of $K^*_2$ as a new
entry to the process is expected to regularize the $\chi^2$-fit to
some degree with a hope that  the double-counting by the duality
between $s$- and $t$-channels should not be significant from the
smallness of the $N^*$ contribution expected.

%--------------------------------------------------
\section*{Acknowledgment}
%--------------------------------------------------
The authors are grateful to S.-i. Nam for fruitful discussions
about the present work. This work was supported by the grant
NRF-2013R1A1A2010504, and by the grant NRF-2016K1A3A7A09005580
from National Research Foundation (NRF) of Korea.

\appendix

\section{Density matrix in $t$-channel helicity frame of $\Lambda^*$}

\begin{figure*}[]
\centering
\includegraphics[width=0.7\hsize,angle=0]{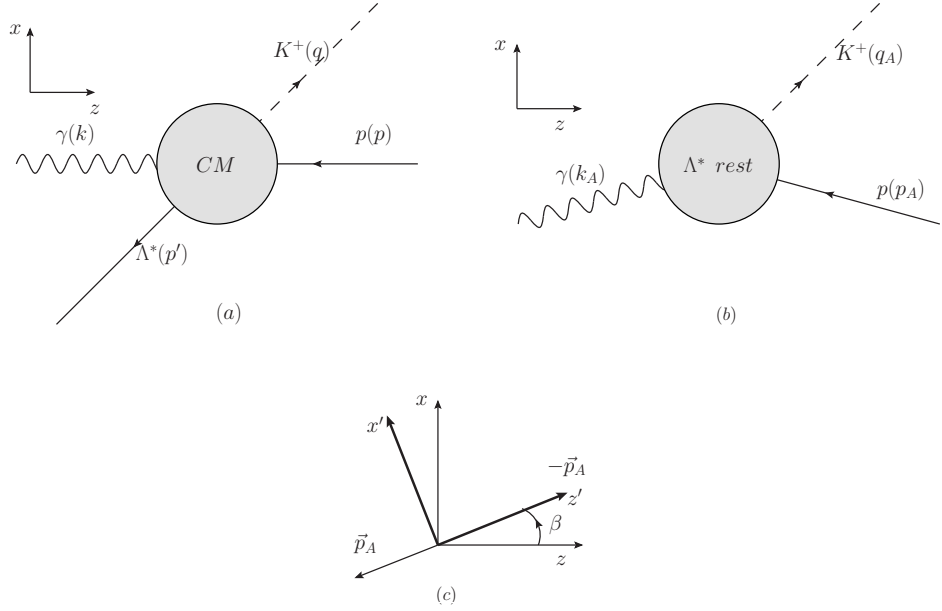}
\caption{Photoproduction in the c.m. frame (a), and in the
$\Lambda^*$ rest frame called as the Adair frame (b). The rotation
from the Adair to the $t$-channel helicity frame of $\Lambda^*$ by
the angle $\beta$ (c).} \label{fig9}
\end{figure*}

We calculate the density matrix elements using the helicity
formalism in the Gottfried-Jackson (G.-J.) frame where the
$\Lambda^*$ is rest. The $z$-axis can be taken  either as the
direction of the incident photon or as the target proton in the
$\Lambda^*$ rest frame conventionally, while the $y$-axis  is
defined to be normal to the production plane, i.e.,
$\hat{y}\propto \hat{\gamma}\times \hat{K}^+ $. In the present
calculation, we take the $z$-axis to be antiparallel to the
direction of the target proton following the Ref. \cite{barrow},
which is  called the $t$-channel helicity frame of $\Lambda^*$. To
work with the Helicity formalism, we perform the Lorentz
transformation of the kinematic variables $k, q, p, p'$ (which are
the respective momenta of the photon, $K^+$, proton, and
$\Lambda^*$) from the center of mass frame to the $\Lambda^*$ rest
frame. By doing this the boosted momenta $k_A, q_A, p_A,p'_A$ are
obtained in the $\Lambda^*$ rest frame, the so-called the Adair
frame. Then, we construct the helicity eigenstate for each
momentum in the Adair frame.

For the target proton with mass $M_p$, we use the helicity
solutions as follows,
\begin{eqnarray}\label{spinor-h1}
\psi(p_A,{1\over2})=N\left(\begin{array}{c}
\cos{\theta\over2}\\
e^{i\phi}\sin{\theta\over2}\\
{|\vec{p}_A|\over E+M_p}\cos{\theta\over2}\\
{|\vec{p}_A|\over E+M_p}e^{i\phi}\sin{\theta\over2}
\end{array}
\right),%\hspace{0cm}
\end{eqnarray}

\begin{eqnarray}\label{spinor-h2}
\psi(p_A,-{1\over2})=N\left(\begin{array}{c}
-\sin{\theta\over2}\\
e^{i\phi}\cos{\theta\over2}\\
{|\vec{p}_A|\over E+M_p}\sin{\theta\over2}\\
-{|\vec{p}_A|\over E+M_p}e^{i\phi}\cos{\theta\over2}
\end{array}
\right),
\end{eqnarray}
where $p_A=(E, |\vec{p}_A|\sin\theta\cos\phi,
|\vec{p}_A|\sin\theta\sin\phi, |\vec{p}_A|\cos\theta)$ is the the
momentum of target proton in the Adair frame and $N$ is the
normalization constant.

For the $\Lambda^*$ resonance, the Rarita-Schwinger field of
spin-3/2 baryon is constructed from the verctor-spinor
representation
\begin{eqnarray}\label{rs3}
&&\psi^\mu(p,{3\over2})=\epsilon^\mu_+(p)u(p,{1\over2})\,,\nonumber\\
&&\psi^\mu(p,{1\over2})=\sqrt{{2\over3}}\epsilon^\mu_0(p)u(p,{1\over2})
+\sqrt{{1\over3}}\epsilon^\mu_+(p)u(p,-{1\over2})\,,\nonumber\\
&&\psi^\mu(p,-{1\over2})=\sqrt{{1\over3}}\epsilon^\mu_-(p)u(p,{1\over2})
+\sqrt{{2\over3}}\epsilon^\mu_0(p)u(p,-{1\over2})\,,\nonumber\\
&&\psi^\mu(p,-{3\over2})=\epsilon^\mu_-(p)u(p,-{1\over2})\,,
\end{eqnarray}
where $p =(M_{\Lambda^*},0,0,0)$, and $M_{\Lambda^*}$ is the mass
of the $\Lambda^*$. For the spinor $u(p)$, the following wave
functions are used

\begin{eqnarray}\label{spinor-r1}
u(p,{1\over2})=\left(\begin{array}{c}
1\\
0\\
0\\
0
\end{array}
\right),%\hspace{0cm}
\end{eqnarray}

\begin{eqnarray}\label{spinor-r2}
u(p,-{1\over2})=\left(\begin{array}{c}
0\\
1\\
0\\
0
\end{array}
\right).
\end{eqnarray}

For the vector $\epsilon^\mu(p)$, the following helicity wave
functions are used
\begin{eqnarray}\label{vector-2}
&&\epsilon^\mu_{\pm}(p)={1\over\sqrt{2}}(0,\,\mp1,-i,0)\,,\\
&&\epsilon^\mu_{0}(p)=(0,0,0,1)\,.
\end{eqnarray}

Once we get the density matrix elements $\rho_A$ in the Adair
frame, we need to perform the coordinate transformation once more
from the Adair to the G.-J. frame to obtain  the $\rho_{GJ}$
there. In other words, to go to the G.-J. frame where the $z$-axis
is antiparallel to the incident proton direction, we need to
rotate the $\rho_A$ about $y$-axis  using the Wigner rotation
matrix, i.e.,
\begin{eqnarray}
\rho_{GJ}=d^\dagger(\beta) \rho_A d(\beta).
\end{eqnarray}
Here the $\beta$ is the angle of the rotation between the incident
proton direction and the $z$-axis in the Adair frame as shown in
Fig. \ref{fig9}. We summarize the procedure discussed above in a
diagrammatic representation in Fig.  \ref{fig9}.

\section{Angular distribution function for $\Lambda^*\to K^- p$}

The angular distribution function $W(\theta,\phi)$ for
$\Lambda^*\to K^- p$ decay measures the decay angle
$(\theta,\phi)$ of $K^-$ in the G.-J. frame (the same angles
$(\theta^*,\phi^*)$ in the text) with respect to the incident
photon polarization;
\begin{eqnarray}
\vec{P}_\gamma=P_\gamma\left(-\cos2\Phi,\,-\sin2\Phi,\,0\right),
\end{eqnarray}
where $\Phi$ is the angle between the polarization vector of the
photon and reaction plane, and
\begin{eqnarray}\label{b2}
W(\theta,\phi,\Phi)=W^0(\theta,\phi)+\sum_{i=1}^3
P_\gamma^i(\Phi)W^i(\theta,\phi)\,.
\end{eqnarray}

The angular distribution function $W^\alpha$ for $\alpha=0, 1,2,3$
is expressed  in terms of the density matrix elements as
\cite{chung}
\begin{widetext}
\begin{eqnarray}
W^\alpha(\theta,\phi)={2J+1\over8\pi}\sum_{\lambda''}D^{(3/2)*}_{\lambda
\lambda''}(\phi,\theta,-\phi)\,
\rho^\alpha_{\lambda\lambda'}\,D^{(3/2)}_{\lambda'\lambda''}(\phi,\theta,-\phi)
\end{eqnarray}
for the decay of spin-${3\over2} \to$ spin-0+spin-${1\over2}$ with
$\lambda''=+{1\over2}$, or $-{1\over2}$, i.e.,
\begin{eqnarray}
W^\alpha(\theta,\phi)={1\over2\pi}\left[ D^{(3/2)*}_{\lambda
{1\over2}}(\phi,\theta,-\phi)\,
\rho^\alpha_{\lambda\lambda'}\,D^{(3/2)}_{\lambda'{1\over2}}(\phi,\theta,-\phi)
+D^{(3/2)*}_{\lambda -{1\over2}}(\phi,\theta,-\phi)\,
\rho^\alpha_{\lambda\lambda'}\,D^{(3/2)}_{\lambda'-{1\over2}}(\phi,\theta,-\phi)
\right]
\end{eqnarray}
with the spin-${3\over2}$ rotation matrix elements given by
\begin{eqnarray}
&&D^{(3/2)}_{\lambda
{1\over2}}=\left(-\sqrt{3}\cos^2{\theta\over2}\sin{\theta\over2}e^{-i\phi},
\,\cos^3{\theta\over2}-2\cos{\theta\over2}\sin^2{\theta\over2},
\,(2\cos^2{\theta\over2}\sin{\theta\over2}-\sin^3{\theta\over2}
)e^{i\phi}, \, \sqrt{3}\cos{\theta\over2}\sin^2{\theta\over2}
e^{2i\phi}\right),\\
&&D^{(3/2)}_{\lambda
-{1\over2}}=\left(\sqrt{3}\cos{\theta\over2}\sin^2{\theta\over2}e^{-2i\phi},
\,(\sin^3{\theta\over2}-2\cos^2{\theta\over2}\sin{\theta\over2})e^{-i\phi},
\,-(2\cos{\theta\over2}\sin^2{\theta\over2}-\cos^3{\theta\over2}),
\, \sqrt{3}\sin{\theta\over2}\cos^2{\theta\over2}
e^{i\phi}\right).\hspace{0.5cm}
\end{eqnarray}
Here, the density matrix elements are
\begin{eqnarray}\label{sdm33}
\rho^\alpha_{\lambda\lambda'}= \left(
\begin{array}{cccc}
\rho_{33} & {\rm Re}\,\rho_{31}+i\,{\rm Im}\,\rho_{31} & {\rm Re}\,\rho_{3-1}+i\,{\rm Im}\, \rho_{3-1}& i\,{\rm Im}\,\rho_{3-3}\\
{\rm Re}\,\rho_{31}-i\,{\rm Im}\,\rho_{31} &  \rho_{11} & i\,{\rm Im}\,\rho_{1-1} &{\rm Re}\,\rho_{3-1}-i\,{\rm Im}\, \rho_{3-1}\\
{\rm Re}\,\rho_{3-1}-i\,{\rm Im}\,\rho_{3-1} & -i\,{\rm
Im}\,\rho_{1-1} & \rho_{11} & -{\rm Re}\,\rho_{31}+i\,{\rm
Im}\,\rho_{31}\\
-i\,{\rm Im}\,\rho_{3-3} & {\rm Re}\,\rho_{3-1}+i\,{\rm
Im}\,\rho_{3-1} & -{\rm Re}\rho_{31}-i\,{\rm Im}\, \rho_{31} &
\rho_{33}
\end{array} \right)\ \ \ \ \ \
\end{eqnarray}
for $\alpha=0,\,1$ with ${\rm Re}\,[\rho_{3-3}]=0={\rm
Re}\,[\rho_{1-1}]$ by hermicity and parity, and
\begin{eqnarray}\label{sdm2}
\rho^2_{\lambda\lambda'}=\left(
\begin{array}{cccc}
\rho_{33} & {\rm Re}\,\rho_{31}+i\,{\rm Im}\,\rho_{31} & {\rm Re}\,\rho_{3-1}+i\,{\rm Im}\,\rho_{3-1}& {\rm Re}\,\rho_{3-3}\\
{\rm Re}\,\rho_{31}-i\,{\rm Im}\,\rho_{31} &  \rho_{11} & {\rm Re}\,\rho_{1-1} &-{\rm Re}\,\rho_{3-1}+i\,{\rm Im}\,\rho_{3-1}\\
{\rm Re}\,\rho_{3-1}-i\,{\rm Im}\,\rho_{3-1} &{\rm Re}\,\rho_{1-1}
& -\rho_{11} & {\rm Re}\,\rho_{31}-i\,{\rm
Im}\,\rho_{31}\\
{\rm Re}\,\rho_{3-3} & -{\rm Re}\,\rho_{3-1}-i\,{\rm
Im}\,\rho_{3-1} & {\rm Re}\rho_{31}+i\,{\rm Im}\, \rho_{31} &
-\rho_{33}
\end{array} \right)\hspace{0.5cm}
\end{eqnarray}
with ${\rm Im}\,[\rho_{3-3}]=0={\rm Im}\,[\rho_{1-1}]$.

Thus, the angular distribution functions for $\alpha=0,1$ and
$\alpha=2$ are given by
\begin{eqnarray}
&&W^\alpha(\theta,\phi)={3\over4\pi}\left[\rho^\alpha_{33}\sin^2\theta
+\rho^\alpha_{11}\left({1\over3}+\cos^2\theta\right)
-{2\over\sqrt{3}}\,{\rm
Re}\,\rho^\alpha_{31}\sin2\theta\cos\phi-{2\over\sqrt{3}}
\,{\rm Re}\,\rho^\alpha_{3-1}\sin^2\theta\cos2\phi\right]\ ,\\
&&W^2(\theta,\phi)={3\over4\pi}\left({2\over\sqrt{3}}\,{\rm
Im}\,\rho^2_{31}\sin2\theta\sin\phi+{2\over\sqrt{3}}\,{\rm
Im}\,\rho^2_{3-1}\sin^2\theta\sin2\phi\right)\ .
\end{eqnarray}

Therefore, according to Eq. (\ref{b2}), the decay angular
distribution in the G.-J. frame is given by \cite{Bingham:1970}
\begin{eqnarray}\label{decay}
&&W(\theta,\phi,\Phi)={3\over4\pi}\Biggl\{
\rho^0_{33}\sin^2\theta+\rho^0_{11}
\left({1\over3}+\cos^2\theta\right) -{2\over\sqrt{3}}{\rm Re}
\left[\rho^0_{31}\cos\phi\sin2\theta+\rho^0_{3-1}\cos2\phi\sin^2\theta\right]\nonumber\\
&&-P_\gamma\cos2\Phi\left[\rho^1_{33}\sin^2\theta+\rho^1_{11}
\left({1\over3}+\cos^2\theta\right) -{2\over\sqrt{3}}{\rm
Re}\left[\rho^1_{31}\cos\phi\sin2\theta+\rho^1_{3-1}\cos2\phi\sin^2\theta\right]\right]\nonumber\\
&&-P_\gamma\sin2\Phi{2\over\sqrt{3}}{\rm
Im}\left[\rho^2_{31}\sin\phi\sin2\theta+
\rho^2_{3-1}\sin2\phi\sin^2\theta\right] \Biggr\}\ .
\end{eqnarray}
\end{widetext}

The beam polarization asymmetry is defined by integrating the $W$
over the $\theta,\,\phi$ angles,
\begin{eqnarray}
\Sigma=-{1\over P_\gamma}{W(\Phi=0)-W(\Phi={\pi\over2})\over
W(\Phi=0)+W(\Phi={\pi\over2})}
\end{eqnarray}
which leads to
\begin{eqnarray}\label{sd1}
\Sigma=2(\rho^1_{33}+\rho^1_{11})\,.
\end{eqnarray}

And in relation with the multikaon process the beam polarization
asymmetry $\Sigma$ with the final kaon measured at the specific
angles $\theta=90^\circ$ and $\phi=90^\circ$ in Eq. (\ref{decay})
is written as  \cite{zhao}
\begin{eqnarray}\label{sd2}
\Sigma^*&=&-{1\over
P_\gamma}{W({\pi\over2},{\pi\over2},\Phi=0)-W({\pi\over2},{\pi\over2},\Phi={\pi\over2})\over
W({\pi\over2},{\pi\over2},\Phi=0)+W({\pi\over2},{\pi\over2},\Phi={\pi\over2})}\,,\nonumber\\
&=&{\rho^1_{33}+{1\over3}\rho^1_{00}+{2\over\sqrt{3}}{\rm
Re}[\rho^1_{3-1}]\over
\rho^0_{33}+{1\over3}\rho^0_{00}+{2\over\sqrt{3}}\,{\rm
Re}[\rho^0_{3-1}]}
\end{eqnarray}
for the measurement at the fixed angle $\theta,\,\phi$.

The $\Sigma$ in Eqs. (\ref{sd1}) is equivalent to the beam
polarization asymmetry
\begin{eqnarray}
\Sigma={d\sigma_y-d\sigma_x\over d\sigma_y+d\sigma_x}
\end{eqnarray}
defined in the center of mass frame. The estimate of the density
matrix elements is in general dependent on the frame chosen for
calculation.

%--------------------------------------------------

\end{document}